\documentclass[superscriptaddress,aps,twocolumn]{revtex4-1}

\pdfoutput=1
\usepackage[utf8]{inputenc}
\usepackage[english]{babel}
\usepackage[T1]{fontenc}
\usepackage{amsmath}
\usepackage{tikz}
\usepackage{lipsum}
\usepackage{physics}
\usepackage{comment}
\usepackage{amsfonts}
\usepackage[toc,page]{appendix}

\usepackage[colorlinks,breaklinks]{hyperref}

\DeclareMathOperator*{\argmax}{arg\,max}

\newcommand{\pytheus}{\textsc{Pytheus}}

\begin{document}
\raggedbottom

\title{Experimental Solutions to the High-Dimensional Mean King's Problem}

\author{Tareq Jaouni}
\email{tjaou104@uottawa.ca}
\affiliation{Nexus for Quantum Technologies, University of Ottawa, K1N 5N6, Ottawa, ON, Canada}

\author{Xiaoqin Gao}
\email{xgao5@uottawa.ca}
\affiliation{Nexus for Quantum Technologies, University of Ottawa, K1N 5N6, Ottawa, ON, Canada}

\author{S\"{o}ren Arlt}
\affiliation{Max Planck Institute for the Science of Light, Erlangen, Germany}

\author{Mario Krenn}
\affiliation{Max Planck Institute for the Science of Light, Erlangen, Germany}

\author{Ebrahim Karimi}
\affiliation{Nexus for Quantum Technologies, University of Ottawa, K1N 5N6, Ottawa, ON, Canada}
\affiliation{Max Planck Institute for the Science of Light, Erlangen, Germany}

\begin{abstract}
In 1987, Vaidman, Aharanov, and Albert put forward a puzzle called the Mean King’s Problem (MKP) that can be solved only by harnessing quantum entanglement. Prime-powered solutions to the problem have been shown to exist, but they have not yet been experimentally realized for any dimension beyond two. We propose a general first-of-its-kind experimental scheme for solving the MKP in prime dimensions ($D$). Our search is guided by the digital discovery framework \pytheus, which finds highly interpretable graph-based representations of quantum optical experimental setups; using it, we find specific solutions and generalize to higher dimensions through human insight. As proof of principle, we present a detailed investigation of our solution for the three-, five-, and seven-dimensional cases. We obtain maximum success probabilities of $72.8 \%$, $45.8\%$, and $34.8 \%$, respectively. We, therefore, posit that our computer-inspired scheme yields solutions that exceed the classical probability (1/$D$) twofold, demonstrating its promise for experimental implementation.
\end{abstract}

\date{\today}
\maketitle

\section*{Introduction}
The Mean King's Problem (MKP) is a quantum puzzle originally posed by Vaidman, Albert, and Aharanov \cite{vaidman_how_1987} that chronicles how one may leverage the properties of quantum entanglement to ascertain the values of the non-commutable Pauli operators $\sigma_x$, $\sigma_y$, and $\sigma_z$ from a spin-1/2 particle. The problem is described as follows: A physicist (Alice) prepares a quantum system to her whims and sends it to a Mean King, who secretly performs projection-valued measurement (PVM) on one of the Pauli spin operators. The King then allows Alice to perform one more experiment, after which He reveals which operator he measured and challenges her to guess his measurement outcome correctly. As we will further elaborate upon in Section \ref{sec:highdim}, the authors reveal that Alice can escape the Mean King's cruelty with a theoretical success probability of 100\% by entangling her particle onto an ancillary particle, which she tends to do in secret, and by choosing to perform a projective valued measurement of her own for her final experiment. 

The MKP has since been adapted to various contexts \cite{botero_mean_2007, yoshida_solution_2015, reimpell_meaner_2007}. Crucially, it has been shown that the author's original trick can also be extended to prime \cite{englert_mean_2001} and prime-powered \cite{hayashi_mean_2005, durt_mutually_2010, revzen_geometrical_2012,  klappenecker_new_2005, bar_2-categorical_2014} dimensions when the Mean King makes His measurement in a maximal set of mutually unbiased bases (MUBs); it has also been shown by \cite{kimura_solution_2006} to exist in composite dimensions (e.g., in $D$=6 and $D$=10) when one allows positive operator-valued measurements (POVMs) to be performed instead. The applicability of the Mean King's Problem has also been considered in Quantum Key Distribution (QKD) \cite{werner_quantum_2009,azuma_interceptresend_2021, bub_secure_2001}. Here, the same protocol can be used to agree on a shared secret key made of a string of Dits (d-dimensional Bits). 

Considering the possible applications of the MKP, it is interesting to implement it experimentally. Quantum optical setups which implement the two-dimensional Mean King's Problem have been proposed \cite{schulz_ascertaining_2003, englert_universal_2001}; In particular, Schulz et al. proposed an experimental realization for the two-Dimensional Mean King's Problem by using an entanglement with two degrees of freedom (DOF), polarization and spatial mode of a single photon. Their experimental setup would have the Mean King Bob imitate a PVM by transforming the input state onto one of several possible choices of eigenstates. It achieved an experimental success probability of $95.6 \%$. An experimental implementation of the tracking the King Problem, a variant to the MKP in which the Mean King does not disclose to Alice His measurement basis, has also been proposed \cite{hu_experimental_2019}, achieving reasonable success probabilities of $81.3 \%$. However, there does not yet exist an experimental proposal for higher dimensions which, particularly within the context of QKD, shows promise due to its robustness to environmental noise \cite{al-amri_front_2021, bouchard_experimental_2018,da_lio_path-encoded_2021}. The difficulty in experimentally implementing the MKP in higher dimensions is due to the final step, in which Alice must measure the non-trivial basis of $D^{2}$ Vaidman-Albert-Aharanov (VAA) states.

This paper illustrates a computer-inspired approach that yields experimental setups for the MKP in arbitrarily high dimensions. We exploit the highly interpretable, graph-theoretical representation of quantum optical experiments \cite{gu_quantum_2019, krenn_quantum_2017}. We determine the correct solution using the AI-based framework \pytheus \ \cite{ruiz-gonzalez_digital_2022} for two and three dimensions, then generalize from these solutions to higher dimensions using human insight. Our experimental solutions can be implemented using linear optical elements and can be realized using any photonic DOF. We demonstrate that our setups' success probabilities significantly exceed classical solutions, demonstrating the quantum advantage of our approach.  

\section*{High-dimensional MKP concept}
\label{sec:highdim}

\begin{figure*}[ht!]  \centering\includegraphics[width=\textwidth]{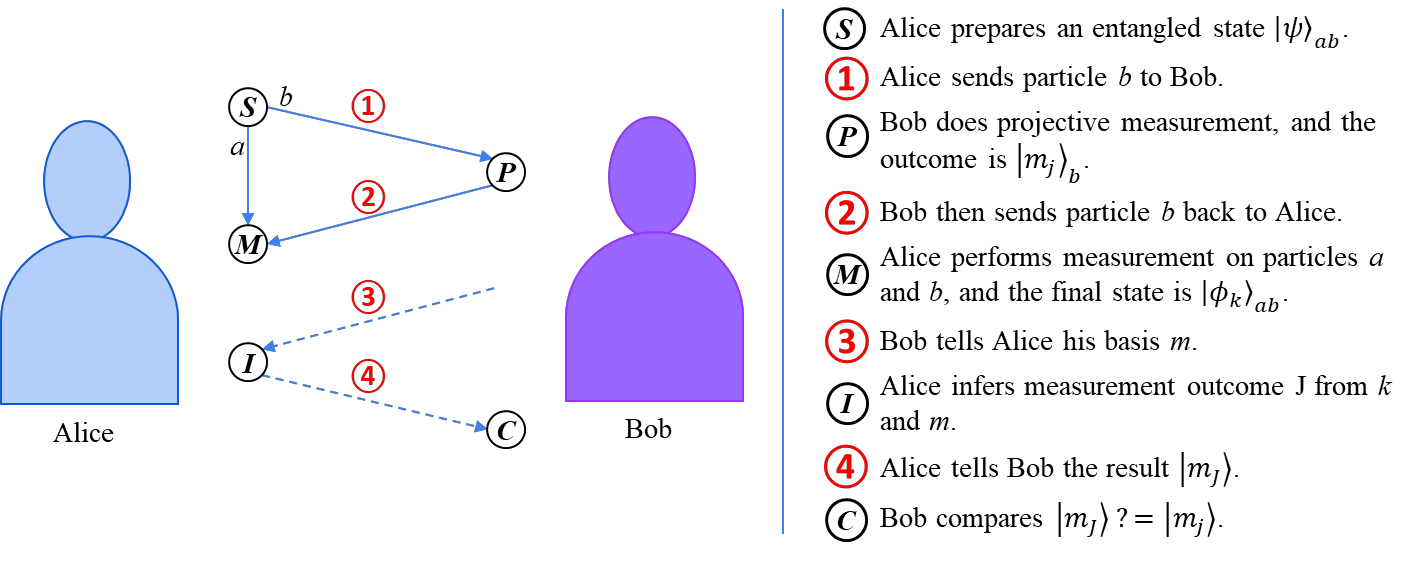}
  \caption{\textbf{Step-by-step process of Alice's solution to the MKP}. She escapes the Mean King's cruelty by entangling and keeping an ancilla secret from the photon she sends to the King. Upon performing a PVM, the King returns Alice's original photon. She then measures both of her photons based on the VAA states, to which the result of her measurement unravels to her the King's measurement result. Over a classical channel, the King reveals to Alice his measurement basis, and Alice, in turn, answers with her guess. }
  \label{fig:figure1}
\end{figure*}

Figure \ref{fig:figure1} shows Alice's protocol for solving the high-dimensional Mean King's problem. We follow the approach of \cite{hayashi_mean_2005}, and confine our study to $D=p > 2$ dimensions, where p is a prime number. The solution may also be adapted to $D=p^{n}$ (or prime-powered) dimensions by working in arithmetic defined by the dimension's corresponding finite field \cite{durt_mutually_2010}.

Alice's stratagem to escape the Mean King's cruelty works as follows. At step $S$ (Send), Alice prepares two entangled photons, labeled $a$ and $b$ for Alice and Bob,  in a generalized Bell state according to 
\begin{equation}
    \ket{\psi}_{ab} = \ket{B_{0,0}} = \frac{1}{\sqrt{D}}\sum_{j=0}^{D-1} \ket{\overline{m_j}}_{a}\ket{m_j}_{b},
\end{equation}
where $\{\ket{m_j}\}^{D-1}_{j}$ is the $j^{th}$ element of the $m^{th}$ MUB in $D$-dimensional Hilbert space, and the conjugate ket, $\ket{\overline{\psi}}$, is defined for any two states $\ket{\psi}$ and $\ket{\phi}$ according to Eq. (3.2) in \cite{durt_mutually_2010}
\begin{equation}
   \ket{\phi}\bra{\psi} = \ket{\overline{\psi}}\ket{\phi} \leftrightarrow \ket{\psi} = \sum_{j=0}^{D-1}{\ket{j}\braket{j}{\psi}^{*}}.
\end{equation}
This convention allows us to choose any MUB in the construction of $\ket{\psi}_{ab}$.

While keeping photon \textit{a} for herself, Alice advances to the Projection ($P$) stage. She sends photon $b$ to the Mean King, who performs a projection-valued measurement (PVM) on the photon in one of $D$+1 possible mutually unbiased bases (MUB) and yields an outcome $j=0,1,\ldots, D-1$. This collapses the overall state to $\ket{m_j}_{a,b} =  \ket{\overline{m_j}}_a\ket{m_j}_b$, where $\ket{m_j}$ refers to the $j^{th}$ eigenstate in the $m^{th}$ MUB.

After having received photon $b$ back from the King, Alice moves to the Measure ($M$) stage: she must now retrodict the King's measurement result $\ket{m_j}_{b}$. To achieve this with complete certainty, Alice performs a simultaneous measurement of both photons based on the VAA states. These are defined as 
\begin{equation}
\ket{\phi}_{k} = -\ket{B_{0,0}} + \frac{1}{\sqrt{D}}\sum_{m=0}^{D}{\ket{m_{f_{k}(m)}}_{a,b}},
\label{(3)}
\end{equation}
where $f_k(m)$ is a mapping function uniquely defined for each VAA state which maps the King's possible measurement bases to Alice's estimate of His outcome. Following \cite{hayashi_mean_2005}, the mapping function can be defined as a constructor for a set of $D+1$ mutually orthogonal Latin squares:
\begin{equation}
f_{k}(m) = \begin{cases}
(m \times i - j)\mod D, \hspace{0.86cm} \text{if} \ m< D, \\\hspace{1.65cm}i \hspace{2.4cm} \text{if} \  m=D,
\end{cases}
\end{equation}
where $(i,j)$ are indices corresponding to an entry in the $m^{th}$ Latin Square. These indices are obtained via the decomposition $k = jD + i$. 

Having completed her measurement, the King now discloses his measurement basis to Alice over a classical channel and awaits Alice's response. The VAA states have the crucial property that 
\begin{equation}
\braket{\phi_k}{m_{j}}_{b} =  \frac{1}{\sqrt{D}} \delta_{j, f_k(m)}.
\end{equation}

\begin{figure*}[ht!]
\centering\includegraphics[width=\textwidth]{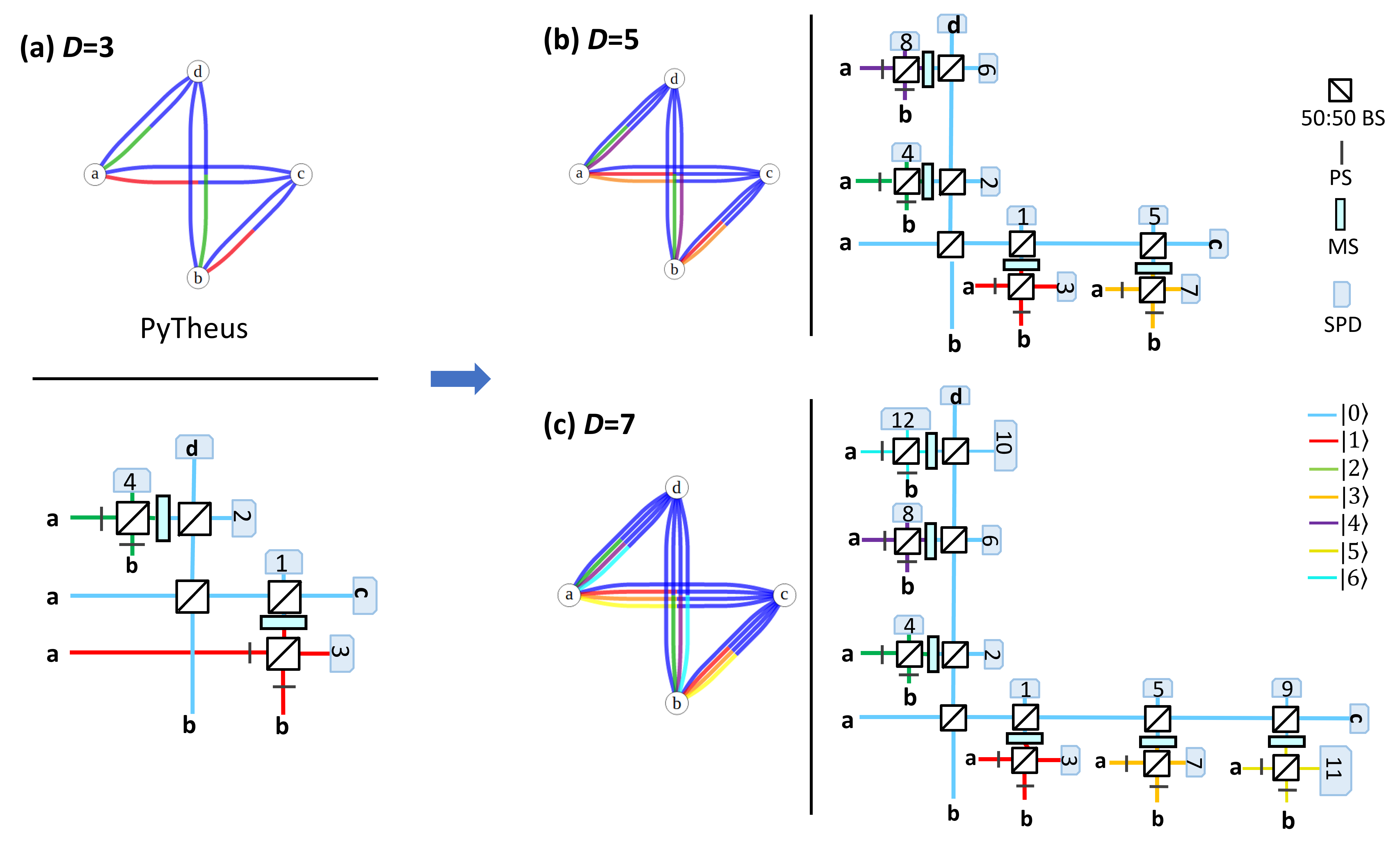}
  \caption{\textbf{High-dimensional MKP Concept.} Quantum Optical Graphs and their corresponding experimental translation for high-dimensional MKP. We show experimental setups corresponding to schemes for \textbf{(a)} three-dimensional, \textbf{(b)} five-dimensional, and  \textbf{(c)} seven-dimensional MKP. We determine the three-dimensional graph-theoretical solution shown in \textbf{(a)} via the digital discovery framework \pytheus. Then, we carry out the experimental translation by creating direct paths from incident photons to detectors. Additional loss detectors, $N_0$ where $N\in\mathbf{N}$, are placed at exit ports of the beamsplitters to provide us with additional information. The three-dimensional solution does not allow simultaneous clicks for higher-order photon modes; we may exploit this idea to ascertain the solutions for arbitrarily high dimensions easily.
  BS: Beam splitter, PS: Phase shifter, MS: Mode sorter, SPD: Single photon detector.
  }
  \label{fig:figure2}
\end{figure*}

In other words, for each of King's measurement results, there exists a disjoint subset of VAA states, $\{ \ket{\phi_{k}}\}_{\ket{m_j}_{b}}$ which encompasses Alice's possible measurement outcomes. Therefore, regardless of which outcome she obtains within this subset, Alice always guesses $\ket{m_{j}}_{b}$, and ends up escaping the Mean King's cruelty with a success probability of 1.

\section*{Work Principle}

We search for experimental realizations of the MKP using an algorithm based on a graph-theoretical representation of quantum linear optical setups \cite{gu_quantum_2019, krenn_quantum_2017}. We ascertain the graph-theoretical solutions' general form using \pytheus  \cite{ruiz-gonzalez_digital_2022}, then translate them into experimental setups using the procedure outlined in section 1 of the Appendix. We conceive our graph-theoretical representations such that one VAA state coincides with two detectors in our setup. Surprisingly, however, we also find it possible to project onto other VAA states by adding additional detectors where a beam splitter joins two paths, and we observe loss, increasing the number of possible two-fold detection events associated with each VAA state.

We then optimize the setup further to project across the entire VAA basis by employing a post-processing optimization algorithm on the phase shifters of our setup. We may compute the success probability  $p_{V}$ that, given the input VAA state $\ket{\phi_k}$, Alice concludes that she measures $\ket{\phi_k}$, using the procedure outlined in Section 2 of the Appendix. We then tune the phase shifters $\varphi = ( \varphi_0, \varphi_1, ... )$ in our setup by employing the  Broyden-Fletcher-Goldfarb-Shanno (BFGS) algorithm \cite{noauthor_numerical_2006} with a loss function defined as

\begin{equation}
    L(\varphi) = -p_{V}(\varphi),
\end{equation}

\begin{table*}[t]
\caption{MKP Success Probability ($p_{M}$) for Different Dimensions}
\begin{tabular}{p{0.2\textwidth}*9l}   
 &&&  \\\toprule
$D$ &  m=0  & m=1  & m=2  & m=3 & m=4 & m=5 & m=6 & m=7 \\ 
3 & 83.3 \% & 62.3 \% & 50.6 \% & 41.8 \% & - & - & - & - \\ 
5 & 53.3 \% &  32.8 \% & 37.6 \% &  36.9 \% & 38.2 \% & 31.2 \% & - & -  \\
7 & 41.8 \% & 24.4 \% & 26.1 \% &  20.8 \% & 23.0 \% & 25.1 \% & 22.8 \% & 27.8 \% \\
 \hline
\end{tabular}
\centering
\label{table: table1}
\end{table*}

We run the algorithm for 20000 iterations with random initialization of the phases to account for the multimodality of the parameter space. We may then quantify our setups' performance regarding the MKP success probability, $p_{M}$, that Alice can successfully retrodict the Mean King's measurement result.

\section*{Results}

\subsection{3D MKP}

We obtain the graph corresponding to the 3D VAA measurement, shown in Fig. \ref{fig:figure2}(a), using  \pytheus \ \cite{ruiz-gonzalez_digital_2022, krenn_conceptual_2021}. 
We choose the third VAA state to be our target.

\begin{align}
        \ket{\phi_3} &= \frac{1}{\sqrt{3}} (\ket{00} + \alpha(\ket{01} + \ket{20})  + \beta(\ket{01} + \ket{20}) \nonumber \\
        &+ \gamma\ket{12} + \delta\ket{21}) 
        + \beta(\ket{12}) + \gamma(\ket{21})), 
\end{align}
where $\omega=e^{-i 2\pi/3}$, $\alpha=(2\omega^2 + \omega)/3$, $\beta=(\omega^2 + 2\omega)/3$, $\gamma=(\omega + 2)/3$ and $\delta = (2 + \omega^2)/3$.     

Fig. \ref{fig:figure2}(a) shows the corresponding experimental setup. The action of the setup on the initial state is given in the operator form $\widehat M(\varphi)$ as 
\begin{equation}
\begin{bmatrix}
a_0 \\
a_1 \\
a_2 \\
b_0 \\
b_1 \\
b_2 
\end{bmatrix}
=
\begin{bmatrix}
\varphi_1 & i\varphi_1 & i\varphi_1 & -\varphi_1 & 0 & 0 \\
-\varphi_3 & 0 & i\varphi_3 & 0 & \varphi_3 & 0 \\
0 & i\varphi_5 & 0 & \varphi_5 & 0 & i\varphi_5\\
i\varphi_2 & \varphi_2 & -\varphi_2 & i\varphi_2 & 0 & 0\\
i\varphi_4 & 0 & \varphi_4 & 0 & i\varphi_4 & 0\\
0 & -\varphi_6 & 0 & i\varphi_6 & 0 & \varphi_6
\end{bmatrix}
\begin{bmatrix}
c_0 \\
d_0 \\
1_0 \\
2_0 \\
3_0 \\
4_0 \\
\end{bmatrix}.
\end{equation}

We report the success probabilities of Alice's measurement for each MUB in Table \ref{table: table1}. After optimization, we compute a success probability of $83\%$ for the m=0 basis, $62.3 \%$ for the m=1 basis, $50.6 \%$ for the m=2 basis and $41.8 \%$ for the m=3 basis. Altogether, this results in an average success probability of $59.5 \%$, which significantly exceeds the classical probability of Alice guessing the measurement result in 3-Dimensions ($\frac{1}{3})$. However, the overall success probability of Alice's MKP measurement depends on the exact set of MUBs chosen by the king. In particular, the maximum success probability of $72.8 \%$ is achieved when Bob only chooses the first two MUBs.

\begin{figure*}[ht!]
\centering
\includegraphics[width=\textwidth]{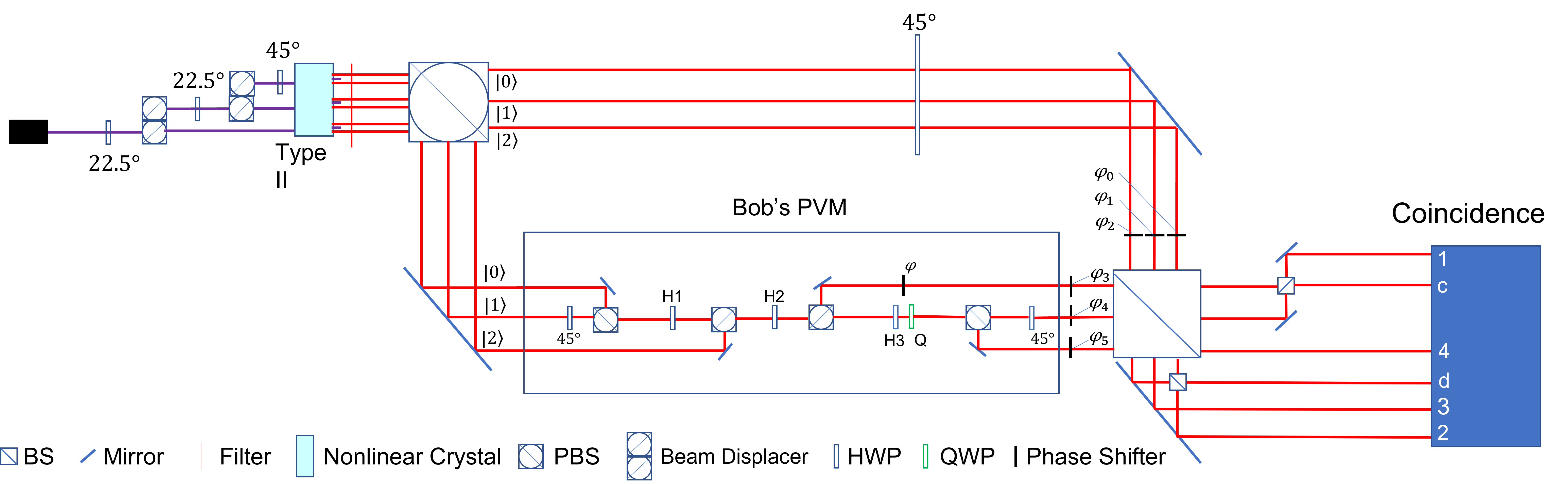}
\caption{\textbf{Experimental Setup.} Example setup that exploits path mode encoding to realize the MKP protocol for three dimensions. Alice exploits the polarization degree of freedom to produce three pump beams that, following SPDC, create three correlated photon pairs. Alice's photon is sent toward the VAA analyzer, while Bob imitates a PVM onto his photon to the eigenstate of his choosing before being sent to the VAA analyzer. PBS is a polarizing beam splitter. H1, H2, and H3 are half waveplates (HWPs), while Q represents a quarter waveplate (QWP). }
\label{fig3:figure3}
\end{figure*}

We provide a possible experimental setup that realizes the three-dimensional MKP in Fig. \ref{fig3:figure3}. Taking inspiration from \cite{PhysRevLett.127.110505}, an ultraviolet (UV) laser is pumped towards a collection of Beam Displacers and half waveplates (HWPs) which projects the incident photons' polarization onto $\ket{H} + \ket{V}$. This results in three beams pumping into a type II $\beta-$Barium Borate (BBO) crystal. It yields three correlated photon pairs, which we may then split through an action of a polarizing beam splitter (PBS); We, therefore, encode the photons' dimensionality using the path mode DOF, while the polarization DOF is used to help achieve certain transformations onto the state. In particular, Bob uses it to achieve his PVM onto an eigenstate of his choosing, using the input configuration provided in \cite{PhysRevLett.123.070505} and the waveplate angle configurations provided in the Appendix. To ensure the photons' indistinguishability, Alice and Bob's polarizations are matched via the action of HWP before being sent to the VAA analyzer described above.

At the cost of increasing complexity, improving upon the average success probability of the VAA measurement further by considering all of the three possible graph-theoretical solutions that we can obtain for the three-dimensional MKP is possible. The $3^2$ VAA states in three dimensions can be partitioned into three disjoint sets: one set contains VAA states with kets $\ket{00}$, but not $\ket{11}$ nor $\ket{22}$. The other sets feature states that satisfy a cyclic permutation of this condition (e.g., $\ket{11}$  but not $\ket{00}$ nor $\ket{22}$). Consequently, Each subset is represented by a different graph-theoretical solution, where simultaneous clicks are conditioned on photons in the same mode. By implementing all three graph-theoretical solutions together, using the recipe described above, we observe a noticeable increase in the success probability. 

\subsection{MKP in Higher Dimensions}

\label{sec:sec3.2}
We make the following inference in order to generalize our computer-inspired solution for the three-dimensional case to higher dimensions: the VAA state given in Eq. \eqref{(3)} has 7 of the nine possible two-qutrit ket states, which determines the set of allowed two-detector click patterns. Crucially, we cannot have a two-fold detector click if both qutrits are of the same higher-order mode (i.e., $\ket{11} or \ket{22}$). In the graph representation, this means that we cannot have perfect matchings realizing the kets $\ket{11}$ or $\ket{22}$, in which each would consist of one edge representing a higher-order mode (1 or 2) between vertices $a$ and one of the detector vertices $c$ or $d$, and another edge of the same mode between vertices $b$ and a different detector vertex. As this is a quality that is shared among the VAA states for D dimensions -- by our construction, the first five VAA states in five-dimensions have $\ket{00}$, but not $\ket{11}$, $\ket{22}$, $\ket{33}$ or $\ket{44}$, for example -- we can build upon the graph-theoretical solution in three-dimensions by adding higher order edges which go from both input mode vertices to \textit{one} output mode detector. It allows us to obtain generalizations of our graph-theoretical solution to arbitrarily high dimensions. 

We demonstrate this idea with a couple of examples: Fig. \ref{fig:figure2}(b) showcases the experimental solution in five dimensions. As before, the input modes $a_i,b_i$ for $i\in \{0,1,2,3,4\}$ are transformed into the detector modes $c_0, d_0$ as well as the loss detector modes $1_0,2_0,...,8_0$ via the operator $\widehat{M(\mathbf{\varphi})}$, with $\mathbf{\varphi} = (\varphi_0, \varphi_2, ..., \varphi_9)$. The MKP success probabilities for the six MUBs in 5D are (in order): $53.3 \%$, $ 32.8 \%$, $37.6 \%$, $36.9 \%$, $38.2 \%$, and $31.2 \%$.  Averaged out over all bases yields a success probability of $38.3 \%$, nearly double the classical probability of $\frac{1}{5}$. We also achieve a maximal success probability of $45.8 \%$ if the King solely chooses to measure on MUBs of $m=0$ and $m=4$.

We iterate upon this idea further by considering the seven-dimensional construction, as shown in Fig. \ref{fig:figure2}(c). For the eight possible MUBs that the Mean King may choose, we report an average success probability of  $26.9 \%$, with a maximal success probability of  $34.8 \%$ if the King chooses the bases m=0 and m=7. This further corroborates our assertion that Alice can double at least her chances of escaping the Mean King's cruelty. We detail the success probabilities of Alice's measurement on the complete set of MUBs in each higher-dimensional case in Table \ref{table: table1}. 

To this point, we have illustrated how to obtain experimental setups for three high-dimensional cases of MKP. By utilizing the abovementioned idea behind generalizing the graph-theoretical solution, we may access ever higher dimensions, allowing us to tap into their increased information capacity and noise robustness benefits. The higher dimensional realizations described before and any other arbitrarily high dimensional variant of the MKP can also be realized using the general experimental setup described in section 4 of the Appendix. 

\section*{Conclusion}

By leveraging graph-theoretical representations of quantum optical experiments, We describe a scheme to formulate quantum optical realizations of Vaidman, Albert, and Aharanov's quantum thought experiment for arbitrarily high dimensions. As a proof of principle, we propose how one may design experimental setups for three-dimensional, five-dimensional, and seven-dimensional cases. Considering every possible choice of MUB by the King, we report maximal success probabilities of $72.8 \%$, $45.8 \%$, and $34. 8\%$ respectively, which exceeds the classical probability of Alice overcoming the Mean King's Problem two-fold. We may posit that our solutions act to double the classical probability ($1/D$), which, owing to their simplicity, demonstrates our solutions' experimental viability. 

Our scheme illustrates another case in which artificial intelligence (AI) and human insight work in tandem \cite{krenn_automated_2016, krenn_computer-inspired_2020}. As the representation of our solutions lends itself to interpretability, we were able to extract insight into the nature of our solutions through the solution to the three-dimensional case and generalize it to solve the more complicated problem of high-dimensional MKP. We hope this work gives further credence to a future in which AI- and human-based intuition work together to discover new science \cite{krenn_scientific_2022}. 

\section*{Acknowledgments}The authors would like to thank Khabat Heshami and Alicia Sit for her valuable discussions, as well as Manuel Ferrer for his help in the current version of the analysis code. This work was supported by the Canada Research Chairs (CRC), the High Throughput and Secure Networks (HTSN) Challenge Program at the National Research Council of Canada, and the Joint Centre for Extreme Photonics (JCEP). 

\section*{References}
\bibliography{arXiv}






\newpage
\clearpage

\appendix

\section{Graph-to-Experiment Translation}
\label{appendix:graphtoexpTranslation}
We convert the graphs featured in this work into experimental setups using the following procedure: We represent incoming photons to the setup as vertices $a,b$ and photonic paths to detectors as vertices $c,d$. We represent paths between incoming and output vertices through weighted, colored multi-edges connecting them. The different edge colorings represent the modes of the photon's DOF, such as its spatial mode or orbital angular momentum. In contrast, the edge weight is phases that we apply to the path using, e.g., phase shifters. We condition the measurement of a state on perfect matchings in the graph -- subsets of edges that have every vertex as endpoints in the graph exactly once -- which correspond to a simultaneous two-fold detector click pattern \cite{gu_quantum_2019, ruiz-gonzalez_digital_2022}.

We systematically translate by determining the perfect matchings corresponding to each ket in our target state. For cases where an input photon has a direct path to both detectors, we place a 50:50 beamsplitter (BS) that splits the beam towards both detectors. In contrast, input photons that exhibit higher order modes are combined, then diverted onto one of the detectors. Higher-order modes are also flattened onto the zero modes using a mode shifter to ensure that the photons are indistinguishable when they arrive at the detectors. 

\section{Success Probability}
\label{sec:appendix}
Towards assigning a metric of the solutions' viability, we attribute the probability that Alice can successfully overcome the Mean King's challenge to our setup. The setup may be modeled as an operator $\widehat{M(\varphi)}$, where $\varphi=(\varphi_0, \varphi_1, ....)$ are phases applied by the tuneable phase shifters in our setup. The effective action of the setup is to transform the input modes $a_i,b_i$, where $i\in(0,1,..D-1)$ corresponds to the $i^{th}$ order mode, into modes corresponding to the output detector modes $c_0, d_0$ as well as additional loss detector modes $1_0, 2_0, ..$. From any input state in the setup, which we may express as sums of monomials of the form $a_ib_j$, we can apply the transformation on the input modes, then expand to yield a polynomial in terms of the detector modes. For example, the output $c_0d_0$ corresponds to the occurrence that detectors c and d click simultaneously. After normalization, the polynomial coefficients give the amplitude of the respective output. Therefore, we obtain a constellation of detector click patterns, along with their corresponding probabilities, associated with each incoming VAA state. This process is achieved computationally with the aid of SymPy \cite{meurer_sympy_2017}.

\begin{table}[t]
\centering
\caption{Waveplate Configurations for Bob's PVM in 3D}
\begin{tabular}{p{0.12\textwidth}*6l}   
 &&&  \\\toprule
State &  H1  & H2 & H3 & Q & $\varphi$ \\ 
(1,0,0) & 22.5° & -22.5° & 0 & 0 & 0   \\ 
(0,1,0) & 22.5° & 22.5° & 0 & 0 & 0   \\ 
(0,0,1) & 22.5° & 22.5° & 45° & 0 & 0  \\ 
(1,1,1) & 22.5° & 0 & 0 & 45° & 0  \\ 
 (1,$\omega$, $\omega^{2}$) & 22.5° & 0 & 60° &  45° & 0  \\
(1,$\omega^{2}$, $\omega$) & 22.5° & 0 & -60° & 45° &   0 \\
(1,$\omega^{2}$, $\omega^{2}$) & 22.5° & 0 & 0 & 45° & -60° \\
(1, $\omega$, 1) & 22.5° & 0  &-60° & 45° & -60°  \\
(1, 1, $\omega$)  & 22.5° & 0 & 60° & 45° & -60°  \\
(1,$\omega$, $\omega$)  & 22.5° & 0 & 0 & 45° & 60° \\
(1,$\omega^{2}$, 1) & 22.5° & 0  & 60° &  45° & 60° \\
(1, 1, $\omega^{2}$) & 22.5° & 0  & -60° & 45° & 60° \\
 \hline
 \label{table2}
\end{tabular}
\end{table}

Alice makes a prior-informed inference to determine the VAA state that she ends up measuring. Given that Alice has measured one of the $\binom{d}{2}$ possible two detectors click patterns $d_{i}$, Alice concludes that she has measured the VAA state that is \textit{most likely to trigger the pattern}. Specifically, for each detector pattern $d_i$, Alice associates to it the VAA state $\argmax{P(\phi_{k}|d_i)}$, where 
\begin{equation}
        P(\phi_k|d_{i})  = \frac{P(d_i|\phi_{j})}{\sum_n^{D^2}P(d_i|\phi_n)},
\end{equation}
where $P(d_i|\phi_n)$ is the probability that the VAA state $\ket{\phi_n}$ triggers detector $d_i$, which is also obtained as the probability amplitude in the VAA state's corresponding polynomial expansion. The overall success probability of Alice's strategy in terms of measuring the correct VAA state is then computed as
\begin{equation}
    p_{V} = \sum_i^{N_{d}}P(d_{i})P(\phi_{MP}|d_{i}), 
\end{equation}
where $N_{d} = \binom{d}{2}$ refers to the number of possible two-detector click patterns with d-detectors in the setup, $P(d_i)$ is the probability that the $i^{th}$ two-detector click pattern occurs, which we may define as 
\begin{equation}
     P(d_i) = \sum_n^{d^2}\frac{P(\phi_{n})}{d^2}, 
\end{equation}
and $P(\phi_{MP}|d_{i})$ is the likelihood that Alice concludes that she has measured $\phi_{MP}$, where $MP = \argmax P(\phi_k|d_i)$. This metric quantifies the setup's ability to distinguish among the $D^2$ VAA states and serves as the loss function for our phase optimization routine. We provide the code along with the data for optimal phases in Code 1 \cite{MKPy}.


\section{Three-Dimensional PVM}
\label{3DPVM}
Table \ref{table2} lists the configurations of the half-and quarter-waveplates required to project onto all states of the MUBs in three dimensions. We note that the state amplitudes are given without normalization as
\begin{equation}
(a, b, c) = a\ket{0}+b\ket{1} + c\ket{2}.
\end{equation}

\label{sec:setup}
\begin{figure*}[t]
\centering
\includegraphics[width=1\textwidth]{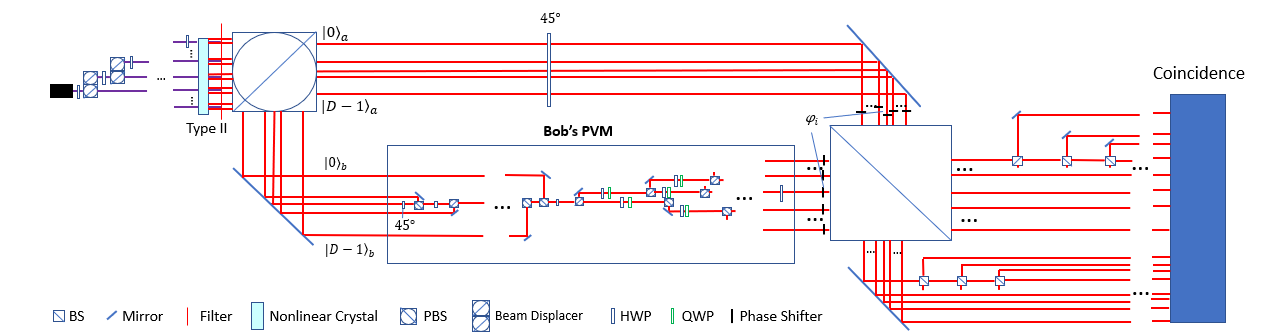}
\caption{\textbf{Experimental setup that realizes MKP in arbitrarily high dimensions.} One obtains the setup by iterating upon the ideas employed in the three-dimensional setup. }
\label{fig:figure4}
\end{figure*}

\section{Experimental Setup in $D$ Dimensions}

We show how we may realize the MKP protocol for higher dimensions in Fig. \ref{fig:figure4}. We iterate upon parts of the experimental setup in the three-dimensional case: For $D$-dimensions, we may introduce $D-1$ BDs to create $D$ spatially separate pump beams and, following Type II SPDC, $D$ correlated photon pairs. Similarly, Bob can introduce repeated iterations of $D-1$ PBSes and $D-1$ polarization controllers to combine and re-separate the photons from one another and project onto any eigenstate of the $D+1$ possible MUBs. As before, Alice's and Bob's photons are made equal in polarity by the action of HWP and then sent toward the VAA analyzer for higher-dimensional setups, as described in section \ref{sec:sec3.2} of the main text.


\end{document}